\documentclass[aps,prl,twocolumn]{revtex4}
\usepackage{graphicx}
\usepackage{amsmath}

\begin{document}

\title{Collective dynamics of complex plasma bilayers}

\author{P Hartmann$^{1,2}$}
\author{Z Donk\'o$^{1,2}$}
\author{G J Kalman$^2$}
\author{S Kyrkos$^3$}
\author{K I Golden$^4$}
\author{M Rosenberg$^5$}

\affiliation{$^1$Research Institute for Solid State Physics and Optics
  of the Hungarian Academy of Sciences, H-1525 Budapest, P. O. Box 49,
  Hungary} 
\affiliation{$^2$Department of Physics, Boston College, 140
  Commonwealth Ave, Chestnut Hill, MA, 02467, USA}
\affiliation{$^3$Department of Chemistry \& Physics, Le Moyne College, 
  1419 Salt Springs Road, Syracuse, NY 13214, USA} 
\affiliation{$^4$Department of Mathematics and Statistics, 
  College of Engineering and Mathematical Sciences, University of
  Vermont, Burlington, VT 05401-1455, USA} 
\affiliation{$^5$Department of Electrical and Computer
  Engineering, University of California San Diego, La Jolla, CA,
  92093, USA}

\begin{abstract}
A classical dusty plasma experiment was performed using two different
dust grain sizes to form a strongly coupled asymmetric bilayer
(two closely spaced interacting monolayers) of two species of charged
dust particles. The observation and analysis of the thermally
excited particle oscillations revealed the collective mode structure
and wave dispersion in this system; in particular the
existence of the theoretically predicted $k=0$ energy (frequency) gap
was verified. Equilibrium molecular dynamics simulations
were performed to emulate the experiment, assuming Yukawa
type inter-particle interaction. The simulations and analytic
calculations based both on lattice summation and on the QLCA approach
are in good agreement with the experimental findings and help
identifying and characterizing the observed phenomena.

\end{abstract}

\maketitle 

Particle bilayers (parallel planes occupied by interacting particles
and separated by a distance comparable to the interparticle distance
within the layers) can be viewed as an intermediate stage between
two-dimensional (2D) and three-dimensional (3D) systems. It is the
interplay of the 3D interaction and the 2D dynamics that creates the
rich new physics predicted and observed in interacting bilayers that
makes these systems interesting in their own right. At the same time,
bilayer configurations are also ubiquitous in widely different
physical systems. Of special importance are those of charged particles
(with like charges: unipolar bilayer, or with opposite charges:
bipolar bilayer). Examples are semiconductor heterostructures
\cite{Seamons2009}, cryogenic traps \cite{Bollinger1998}, overdamped
system of lipid membranes \cite{Watkins2009}, interfacial
superconductors \cite{Yuzephovich2008}, etc.

From the theoretical point of view, during the past three decades
unipolar layered systems were studied in the weak coupling limit by
means of analytic calculations \cite{DasSarma1981,Vilk1985} and in the
strongly coupled regime by semi-analytic lattice calculations
\cite{Goldoni1996}, approximate liquid state calculations
\cite{Kalman1993} and computer simulations
\cite{Schweigert1999,Donko2001,Donko2003,Lowen2003,Ranganathan2004}. More
recently, bipolar (electron--hole) bilayers have been the focus of
intense computer simulation efforts, both in the classical
\cite{Hartmann2005,Kalman2007,Ranganathan2007} and in the quantum
regimes \cite{Filinov2006,DePalo2008}, where features such as
structural phase transitions, bound dipole formation \cite{Astra2007},
etc. were detected.

Very recently a seminal observation by Hyde {\it et.al.}
\cite{Hyde2008} has led to the realization that strongly coupled
unipolar bilayers can be created in laboratory complex (dusty) plasma
environments. This can be accomplished by using a mixture of two
differently sized grains: in view of their necessarily different
charge to mass ($Z/m$) ratios, the two species would settle at
different equilibrium heights in the plasma sheath, as governed by the
local balance of gravitational and electric forces. Thus this novel
type of bilayer would be, in contrast to most of the previously
observed ones, a binary bilayer with hitherto unexplored features. Its
structural properties were already reported in
\cite{Hyde2008}. Subsequent computer simulations predicting collective
excitations (wave propagation) and their dispersion were carried out
in recent years both for Coulomb and Yukawa type isotropic
interactions \cite{Hyde2006,Ranganathan2008}.

In this Letter we report on the experimental investigation of the
collective dynamical properties of a binary dusty plasma system. Our
results constitute the first observation of the mode spectrum of a
strongly coupled (liquid or solid) bilayer. Earlier results
\cite{Kainth2000} were restricted to the weakly coupled state, where
all collective modes exhibit an acoustic behavior. We confirm the
predicted \cite{Kalman1993,Goldoni1996} benchmark of the strongly
coupled mode structure, the development of optic modes with a
wave-number $k=0$ ``energy (frequency) gap''.

Our dusty plasma experiments have been carried out in a custom
designed vacuum chamber with an inner diameter of 25 cm and height of
18 cm. The lower, powered, 18~cm diameter, flat, horizontal, stainless
steel electrode faces the upper, ring shaped, grounded aluminum
electrode with an inner diameter of 15~cm at a height of
13~cm. Experiments have been performed in (4.6 purity) argon gas
discharge at a pressure $p = 0.8 \pm 0.05$ Pa, in a steady gas flow of
$\sim 0.01$ sccm, with 13.56~MHz radio frequency excitation of $\sim
5$~W power. In the experiment melamine-formaldehyde micro-spheres with
diameters $d_1 = 3.63 \pm 0.06 {\rm \mu m}$ and $d_2 = 4.38 \pm 0.06
{\rm \mu m}$ are used. For illumination we apply a 200 mW, 532 nm
dpss-laser. Our CCD camera has a resolution of 1.4 Megapixels and runs
at 29.54 frames per second acquisition rate. Particle masses are:
$m_1=3.8~10^{-14}$ kg and $m_2=6.6~10^{-14}$ kg.

During the evaluation of the raw images (typically over 60000
per experiment) identification and position measurement of the
particles is performed using the method described in
\cite{Goree2007}. The identification of the light and heavy (smaller
and larger) particles is based on their scattered intensities, which
is in this size domain proportional to the square of the
diameter. After tracing the particles motion from frame to frame we
obtain the positions and velocities of each particle as a function of
time. The layer separation was measured by simply rotating the camera
and illuminating setup and taking images through the side
window. After calibrating with a size standard we find the following
average structural parameters: number of particles: $N_1=682$,
$N_2=636$; the Wigner-Seitz radius calculated from the total number
of the observed particles ($N$) and the area ($A$) of the field of
view: $a=\sqrt{A/\pi N}=0.243$~mm (in the following distances
appear normalized to $a$); layer separation $\bar{d}=d/a=0.43$.

\begin{figure}[ht]
\centering
\includegraphics[width=3.0 in]{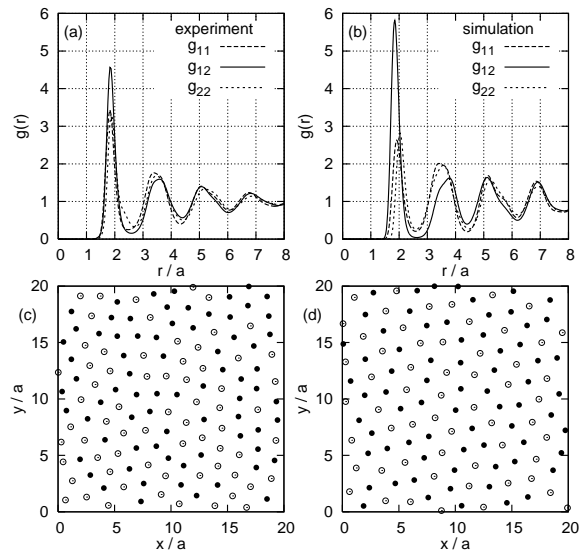}
\caption{Pair-distribution functions and particle snapshots obtained
from the experiment (a,c) and the MD simulation (b,d). Indexes and
symbol types label the layers [open: layer~1, filled: layer~2].}
\label{fig_pcf}
\end{figure}

Based on the particle positions the $g(r)$ pair distribution function
is obtained and compared to molecular-dynamics (MD) simulation
results, (which was carried out by using the same input parameters as
in the experiment), as shown in Fig. \ref{fig_pcf}. Comparing
experiment and simulation, one can conclude, that both capture
qualitatively identical features: order, shape and position of the
peaks agree satisfactorily. Differences in the amplitude and decay
rate are due to friction and density gradients due to the finite and
confined nature of the experimental system. In the simulation infinite
system size (periodic boundaries) and no friction are assumed. Also,
the layer assignment procedure during the data processing has some
degree of uncertainty. Peak positions are consistent with the
underlying hexagonal lattice structure, in agreement with theoretical
results in \cite{Goldoni1996,Lowen2003}.

Visual observation of the recorded images confirms that most of the
system is in a hexagonal configuration, sometimes small domains with
rhombic and square unit cells were also found. Although the ground
state configuration for a bilayer is a regular staggered rectangular
lattice, a large degree of substitutional disorder \cite{KalmanCM},
due to the small inter-layer separation and the finite temperature,
was observed, as shown in the particle snapshots in
Fig.~\ref{fig_pcf}.

In the MD simulation Yukawa type pair-interaction $\Phi_{ij}
=(e^2Z_iZ_j/4\pi\varepsilon_0)\exp[-r_{ij}/\lambda_D]/r_{ij}$, where
$r_{ij}$ is the three-dimensional distance between particles $i$ and
$j$, is assumed using $\sim 4000$ particles with periodic boundary
conditions. Since the interlayer separation is of the order of $1/4
\lambda_D$, the attractive force between grains in the two layers due
to ion focusing or wakefield effects are assumed to be small, since
these effects occur generally at larger separations (see
e.g. \cite{Lampe00}). The neglect of such attractive forces
is further bolstered by the experimental observation that the grains
in the two layers do not form strings, but rather assume a staggered
structure. Particle densities, masses and layer separation are taken
from the experiment, $\kappa = a/\lambda_D = 0.5$ and $Z_1=2550$,
$Z_2=(d_2/d_1)Z_1 \approx 3080$ are further assumed; this parameter
set was found in lattice calculations to best reproduce experimental
dispersions.

\begin{figure}[ht]
\centering
\includegraphics[width=3.0 in]{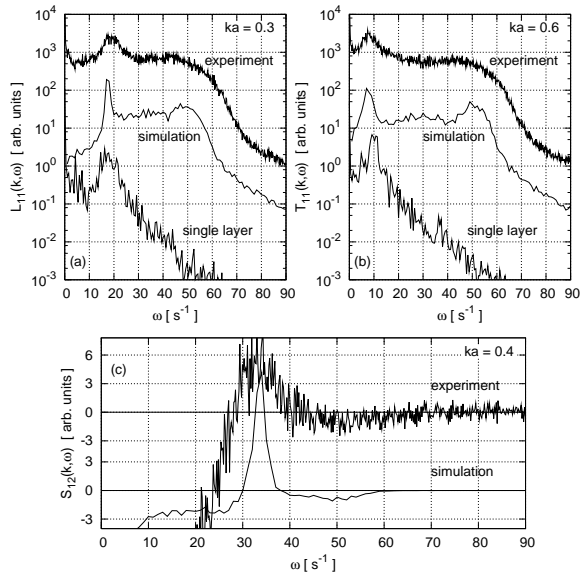}
\caption{Sample spectra illustrating the principal features of
(a) $L_{11}$ longitudinal, (b) $T_{11}$ transverse current
fluctuations and (c) $S_{12}$ inter-layer density fluctuations. The
one-layer spectra from the experiment and MD simulations are shown
and compared to experimental single-layer spectra obtained in a
separate measurement. The inter-layer spectra (experiment and MD
simulation) show a positive and a negative peak representing the
in-phase and out-of phase oscillations, respectively. The vertical
scale is shifted for clarity.}
\label{fig_spec}
\end{figure}

In the experiment we observe a pressure jump of about 38\% when
switching on the discharge. Taking into account the active and the
total volume of the vacuum system, this results in an $\approx$ 85\%
average temperature increase in the discharge region, resulting in
$T_p \approx 550 K$. Measuring the dust particle velocities gives the
direct kinetic energy, resulting in $T_{v1} \approx 400 K$ and $T_{v2}
\approx 500 K$ for the two layers, and the Coulomb coupling parameters
[defined as $\Gamma_m = (e^2/4\pi\varepsilon_0) (Z_m^2/k_{\rm B}T_m
a_m)$]: $\Gamma_1 \approx 800$ and $\Gamma_1 \approx 900$.  To obtain
dynamical information on the system's collective excitations we use
the method based on the Fourier transform of the microscopic density
and current fluctuations, already successfully applied in MD
simulations, see e.g. \cite{Donko2003,Hansen}. Knowing the particle
positions vs. time, first we calculate the microscopic densities, as
well as longitudinal and transverse currents: $\rho^{(m)}({\bf
k},t)=\sum_j \exp(-i{\bf k}\cdot{\bf r}_j)$, $\lambda^{(m)}({\bf
k},t)=\sum_j v_{j,\parallel}\exp(-i{\bf k}\cdot{\bf r}_j)$,
$\tau^{(m)}({\bf k},t)=\sum_j v_{j,\perp}\exp(-i{\bf k}\cdot{\bf
r}_j)$ for the two layers $(m=1,2)$. From the Fourier transforms of
$\rho({\bf k},t) \rightarrow \rho({\bf k},\omega)$, $\lambda
\rightarrow \lambda({\bf k},\omega)$, and $\tau \rightarrow \tau({\bf
k},\omega)$ one can calculate the power spectra, e.g. $S_{m,n}({\bf
k},\omega)\propto\langle \rho^{(m)}(-{\bf k},-\omega)\rho^{(n)}({\bf
k},\omega)\rangle$, averaging is over the time-slices available. In
the present case of an asymmetric bilayer the labeling of the modes is
not possible in a simple way, as it is for symmetric bilayers
\cite{Donko2003}, where the two ``+'' and ``--'' polarizations clearly
separate and correspond to in-, and out-of-phase
oscillations. Therefore, in the following we do not label the modes,
only indicate in which spectra they appear as peaks. For an analysis
of the mode structure one has to examine the spectra individually and
identify the peak positions. Sample spectra are shown in
Fig.~\ref{fig_spec} to illustrate the fundamental features in more
detail. Plotted are results of the bilayer experiment and of the MD
simulation; also to serve as reference standard, results of a
separately performed single layer experiment carried out in the same
experimental setup and under the same conditions except for using only
one particle species (with diameter $4.38 \pm 0.06 {\rm \mu m}$, the
total particle number in the field-of-view was $N=1945$, resulting in
higher density, thus higher nominal plasma frequency as in the bilayer
case).

In Fig. \ref{fig_spec}(a) and (b) one can observe both in the
experiment and in the simulation the presence of a strong
primary peak at lower frequencies and a shoulder-like, weak and wide
feature at higher ($\omega \approx 50 s^{-1}$) frequencies. The
strength of this high-frequency peak becomes more obvious in
comparison with the single layer situation, where at the same
frequency the spectral power has already dropped two orders of
magnitude, while it has the same slope as seen in the bilayer
spectrum at higher frequencies ($\omega > 70
s^{-1}$). Figure \ref{fig_spec}(c) shows the inter-layer density
fluctuation spectrum $S_{12}$ at $ka=0.4$ from the experiment and the
MD simulation. The novel feature here is the appearance of a negative
peak indicating an out-of-phase oscillation at the higher frequency,
coinciding with the ``shoulder'' in the $L_{11}$ and $T_{11}$
spectra. As it is expected, all spectral peaks are more pronounced
and sharper in the simulation, where frictional damping and
disturbing effect of the finite confinement are absent.

We compare our observational results with the mode
dispersion calculated through two theoretical models. The first model
is a perfect staggered rectangular lattice, with phonon propagation
along the two principal axes. The lattice calculation is based on
the formalism used in \cite{Goldoni1996} for electronic bilayers,
adopted to the asymmetric Yukawa system. The second model is a
completely disordered solid bilayer, with the mode dispersion
calculated in the Quasi-Localized Charge Approximation (QLCA)
formalism. The QLCA formalism was adapted to the binary Yukawa
bilayer system (based on \cite{QLCA,Donko2009}) and the dispersions
for the 4 modes were calculated with the input of the experimental
pair-correlation functions. It should be noted that while for
symmetric configurations the (in general) 4x4 dynamical matrix can be
decomposed into two 2x2 matrices, resulting in a clear separation of
the collective modes into longitudinal vs. transverse and in-phase
vs. out-of-phase polarizations, in the present case, this kind of
mode separation, in general, is not possible. It can be done only
along the directions of the principal axes of the crystal, or in an
approximation where the system is assumed to be isotropic as in the
QLCA description.

\begin{figure}[ht]
\centering
\includegraphics[width=3.4 in]{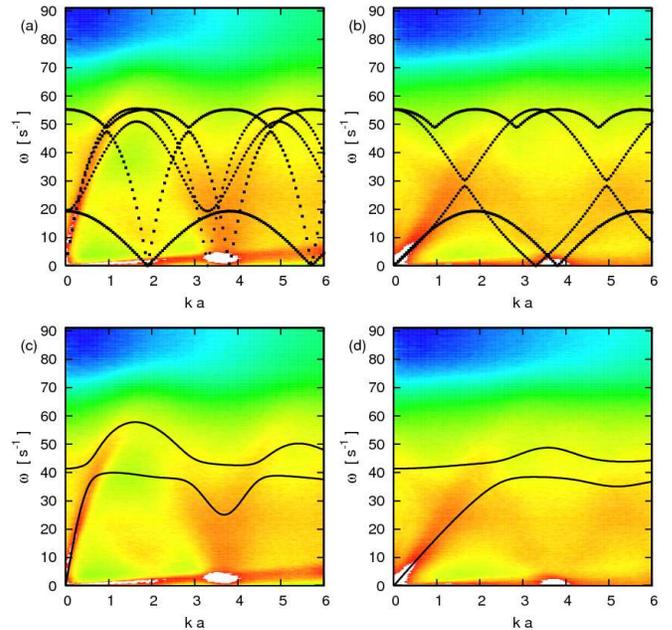}
\caption{(color online) $L_{11}$ one-layer longitudinal (a,c) and
$T_{11}$ one-layer transverse (b,d) current fluctuation
spectra. Peak positions (hot colors) mark the dispersion
$\omega(k)$. Black symbols in (a,b) are frequencies form lattice
summation calculations including the two principal lattice
directions. Lines in (c,d) are the corresponding QLCA
dispersions.}
\label{fig_2D}
\end{figure}

Figure \ref{fig_2D} shows an example of the longitudinal and
transverse current fluctuation spectra in the upper layer, color coded
in the wave-number ($k a$) / frequency ($\omega$) plane. Overlayed are a
selected set of lattice dispersions (panels [a] and [b]), and the QLCA
mode dispersions (panels [c] and [d]).  An additional, very low
frequency, linear dispersion can be seen, which is most likely the
fingerprint of an unavoidable net motion of the ensemble.

\begin{figure}[ht]
\centering
\includegraphics[width=3.0 in]{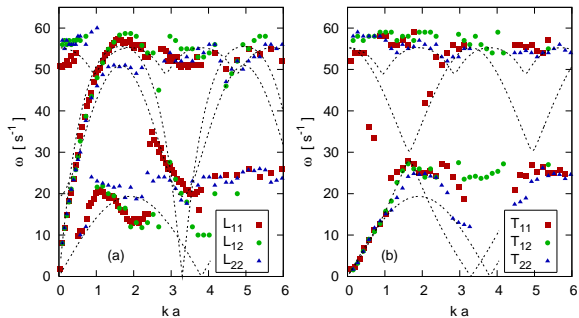}
\caption{(color online) Symbols: peaks identified in the
longitudinal (a) and transverse (b) experimental spectra. Lines:
selected lattice dispersions with $x$ and $y$
polarizations.}
\label{fig_disp}
\end{figure}

Figure \ref{fig_disp} shows the more detailed dispersion properties of
the dusty plasma bilayer together with lattice calculation
results. The modes are not labeled, for reasons discussed above, only
their sources (the specrta in which the peak was found) are indicated.

In Figs. \ref{fig_2D} and \ref{fig_disp} one can observe the expected
acoustic modes similar to those in 2D Yukawa layers, and an optical
excitation with finite and nearly constant frequency, even at very low
wave-numbers. This latter mode is identified as the so called energy
(frequency) gap, studied in great detail in earlier simulations and
theoretically predicted through lattice calculations
\cite{Goldoni1996} and for strongly coupled liquids through the QLCA
approach \cite{Donko2003,Kalman1993}. The QLCA model
provides the value of the gap frequency as

\begin{equation}\label{eq:gap}
\omega_{\rm gap}^2
=\frac{\omega_1^2}{2}\left(\frac{Z_2n_2}{Z_1n_1}+\frac{Z_2m_1}{Z_1m_2}
\right)\int_0^\infty {\cal F}(r)g_{12}(r) {\rm d}r,
\end{equation}
where $\omega_1^2=(e^2Z_1^2n_1)/ (2\varepsilon_0 m_1
\sqrt{a_1a_2})$, the kernel ${\cal F}(r) = z^{-3}r
{\rm e}^{-w}[(1+w)(1-3d^2/z^2)+w^2]$,
$z=\sqrt{r^2+d^2}$, and $w=z/\lambda_{\rm D}$.

In view of the prevailing local lattice structure in coexistence with
a high degree of disorder it is not a priori clear, which of the two
models (lattice vs. QLCA) should provide a better description of the
mode structure. Inspection of Fig. \ref{fig_2D} shows that for low $k$
values the acoustic portions of the low frequency modes are equally
well described by either model. For higher $k$ values the repeated
Brillouin zone structure is clearly visible, indicating the
superiority of the lattice model. For the high frequency optic mode
the QLCA predicts a single ``frequency gap'' at $k=0$ with
Eq.(\ref{eq:gap}): $\omega_{\rm gap}^{\rm exp}=41.4 s^{-1}$ and
$\omega_{\rm gap}^{\rm MD}=43.5 s^{-1}$ calculated with the input of
experimental and simulation $g_{12}(r)$ data. This value and the
high-$k$ tapering off of the optic modes seem to be more along the
line of the QLCA description.

We can conclude, that our dusty plasma experiment using two
different dust sizes created a strongly coupled bilayer system that
can be well approximated by a unipolar binary Yukawa bilayer
model. This model has served as the basis for MD simulations, lattice
calculations, and a QLCA calculation. All these approaches are in
good agreement with the experiment, and verify the presence of an
optical collective mode characterized by a finite energy (frequency)
gap at $k=0$ wave-number, distinguishing the strongly coupled bilayer
from a weakly coupled one, where all the modes have an acoustic
character.

Work partially supported by NSF Grants PHY-0813153,
PHY-0812956, DOE Grants DE-FG02-03ER54716, DE-FG02-04ER54804 and
OTKA-K-77653, OTKA-PD-75113, MTA-NSF/102. This paper was supported by
the Janos Bolyai Research Scholarship of the Hungarian Academy
of Sciences. Experimantal setup was partially donated by the
MOM-Szerviz Kft, and assembled by J. Forg\'acs,
J. T\'oth, and Gy. Cs\'asz\'ar.


\begin{thebibliography}{1}

\bibitem{Seamons2009}
  J. A. Seamons {\it et.al.}, %C. P. Morath, J. L. Reno, and M. P. Lilly, 
  {\it Phys. Rev. Lett.} {\bf 102}, 026804 (2009).

\bibitem{Bollinger1998} 
  T. B. Mitchell {\it et.al.}, 
  % J. J. Bollinger, D. H. E. Dubin, X.-P. Huang, W. M. Itano, and R. H. Baughman,
  {\it Science} {\bf 282}, 1290 (1998).

\bibitem{Watkins2009}
  E. B. Watkins {\it et.al.},
  % C. E. Miller, D. J. Mulder, T. L. Kuhl, and J. Majewski, 
  {\it Phys. Rev. Lett.} {\bf 102}, 238101 (2009).
  
\bibitem{Yuzephovich2008} 
  O. I. Yuzephovich {\it et.al.}, {\it Low
  Temp. Phys.} {\bf 34}, 985 (2008).

\bibitem{DasSarma1981} %5
  S. Das Sarma and A. Madhukar,
  {\it Phys. Rev. B} {\bf 23}, 805 (1981);
  for more references, see those in Ref. \cite{Kalman1993}(c)

\bibitem{Vilk1985}
  Yu. M. Vil'k and Yu. P. Monarkha, {\it Soviet Journal of Low
  Temperature Physics} {\bf 11}, 535 (1985).

\bibitem{Goldoni1996}
  G. Goldoni and F. M. Peeters,
  {\it Phys. Rev. B} {\bf 53}, 4591 (1996).

\bibitem{Kalman1993} 
  (a) K. I. Golden and G. Kalman, {\it
  Phys. Stat. Sol. B} {\bf 180}, 533 (1993); 
  (b) G. J. Kalman, Y. Ren,
  and K. I. Golden, {\it Phys. Rev. B} {\bf 50}, 2031 (1994); 
  (c) G. J. Kalman, V. Valtchinov, and K. I. Golden, {\it
  Phys. Rev. Lett.} {\bf 82}, 3124 (1999).

\bibitem{Schweigert1999}
  I. V. Schweigert {\it et.al.}, %V. A. Schweigert, and F. M. Peeters, 
  {\it Phys. Rev. Lett.} {\bf 82}, 5293 (1999).

\bibitem{Donko2001} %10
  Z. Donk\'o and G. J. Kalman,
  {\it Phys. Rev. E} {\bf 63}, 061504 (2001).

\bibitem{Donko2003}
  Z. Donk\'o et.al, % G. J. Kalman, P. Hartmann, K. I. Golden, and K. Kutasi,
  {\it Phys. Rev. Lett.} {\bf 90}, 226804 (2003); 
  Z. Donk\'o et.al, % P. Hartmann, G. J. Kalman, and K. I. Golden, 
  {\it J. Phys. A: Math. Gen.} {\bf 36} 5877 (2003).

\bibitem{Lowen2003} 
  R. Messina and H. L\"owen,
  {\it Phys. Rev. Lett.} {\bf 91}, 146101 (2003).

\bibitem{Ranganathan2004} 
  S. Ranganathan and R. E. Johnson,
  {\it Phys. Rev. B} {\bf 69}, 085310 (2004).

\bibitem{Hartmann2005}
  P. Hartmann {\it et.al.}, %Z. Donk\'o, and G. J. Kalman,
  {\it Europhys. Lett.} {\bf 72}, 396 (2005).

\bibitem{Kalman2007} %15
  G. J. Kalman {\it et.al.}, %P. Hartmann, Z. Donk\'o, and K. I. Golden,
  {\it Phys. Rev. Lett.} {\bf 98}, 236801 (2007).

\bibitem{Ranganathan2007}
  S. Ranganathan and R. E. Johnson,
  {\it Phys. Rev. B} {\bf 75}, 155314 (2007).

\bibitem{Filinov2006}
  A. Filinov {\it et.al.}, %P. Ludwig, Y. E. Lozovik, and M.  Bonitz,
  {\it J. Phys.: Conf. Ser.} {\bf 35}, 197 (2006).

\bibitem{DePalo2008}
  S. De Palo {\it et.al.}, %F. Rapisarda, and G. Senatore, 
  {\it Phys. Rev Lett.} {\bf 88}, 206401 (2002).

\bibitem{Astra2007}
  G. E. Astrakharchik, J. Boronat, I. L. Kurbakov, and Yu. E. Lozovik, 
  {\it Phys. Rev. Lett.} {\bf 98}, 060405 (2007).

\bibitem{Hyde2008} %20
  B. Smith, T. Hyde, L. Matthews, J. Reay, M. Cook, and J. Schmoke,
  {\it Advances in Space Research} {\bf 41}, 1509 (2008).

\bibitem{Hyde2006} 
  L. S. Matthews , K. Qiao, T. W. Hyde,
  {\it Advances in Space Research} {\bf 38}, 2564 (2006).

\bibitem{Ranganathan2008} 
  S. Ranganathan and R. E. Johnson,
  {\it Phys. Rev. B} {\bf 78}, 195323 (2008).

\bibitem{Kainth2000} 
  D. S. Kainth {\it et.al.}, {\it J. Phys. Cond. Matter} {\bf 12}, 439 (2000).

\bibitem{Goree2007}
  Y. Feng {\it et.al.}, % J. Goree, and Bin Liu,
  {\it Rev. Sci. Instr.} {\bf 78}, 053704 (2007).
  
\bibitem{KalmanCM} %25
   G.J. Kalman et.al, 
   {\it Condensed Matter Theories} {\bf 13}, 203 (1997).  

\bibitem{Lampe00}
  M. Lampe {\it et.al.}, %G. Joyce, G. Ganguli, and V. Gavrishchaka, 
  {\it Phys. Plasmas} {\bf  7}, 3851 (2000).

\bibitem{Hansen}
  J. P. Hansen {\it et.al.},% I. R. McDonald, and E. L. Pollock, 
  {\it Phys. Rev. A} {\bf 11}, 1025 (1975).

\bibitem{QLCA}
  K.I. Golden and G.J. Kalman,
  {\it Phys. Plasmas} {\bf  7}, 14 (2000).

\bibitem{Donko2009}
  Z. Donk\'o {\it et.al.}, {\it Book of Abstracts}, 
  12th Workshop on the Physics of Dusty Plasmas, 
  Boulder, Colorado, USA, May 17 - 20, (2009).

\end{thebibliography}
\end{document}